\documentclass[sigconf,screen,nonacm]{acmart}

\newcommand{\systemname}{EarStreAM}

\usepackage{balance}
\usepackage{soul}
\usepackage{balance}

\begin{document}

\title{\systemname{}: A Closed-Loop Earable System for Personalized Stress-Adaptive Meditation}


\author{Jonas Hummel}
\email{jonas.hummel@kit.edu}
\orcid{0009-0005-8563-6175}
\affiliation{%
  \institution{Karlsruhe Institute of Technology}
  \city{Karlsruhe}
  \country{Germany}
}

\author{Luisa Faust}
\email{ufwim@student.kit.edu}
\orcid{0009-0001-7099-7366}
\affiliation{%
  \institution{Karlsruhe Institute of Technology}
  \city{Karlsruhe}
  \country{Germany}
}

\author{Elias M\"{u}ller}
\email{udjks@student.kit.edu}
\orcid{0009-0003-8901-1450}
\affiliation{%
  \institution{Karlsruhe Institute of Technology}
  \city{Karlsruhe}
  \country{Germany}
}

\author{Eva Bertog}
\email{udyft@student.kit.edu}
\orcid{0009-0007-3087-8307}
\affiliation{%
  \institution{Karlsruhe Institute of Technology}
  \city{Karlsruhe}
  \country{Germany}
}

\author{Valeria Zitz}
\email{valeria.zitz@kit.edu}
\orcid{0009-0004-1158-861X}
\affiliation{%
  \institution{Karlsruhe Institute of Technology}
  \city{Karlsruhe}
  \country{Germany}
}

\author{Marius Johannes Prill}
\email{uwptv@student.kit.edu}
\orcid{0009-0007-3152-4383}
\affiliation{%
  \institution{Karlsruhe Institute of Technology}
  \city{Karlsruhe}
  \country{Germany}
}

\author{Luca L. Bennardo}
\email{luca.bennardo@kit.edu}
\orcid{0000-0001-5046-2260}
\affiliation{%
  \institution{Karlsruhe Institute of Technology}
  \city{Karlsruhe}
  \country{Germany}
}
\author{Luisa Weber}
\email{luisa.weber@kit.edu}
\orcid{0009-0007-7644-0909}
\affiliation{%
  \institution{Karlsruhe Institute of Technology}
  \city{Karlsruhe}
  \country{Germany}
}

\author{Tobias R{\"o}ddiger}
\email{tobias.roeddiger@ipai-foundation.ai}
\orcid{0000-0002-4718-9280}
\affiliation{%
  \institution{IPAI Foundation gGmbH}
  \city{Heilbronn}
  \country{Germany}
}

\author{Michael Beigl}
\email{michael.beigl@kit.edu}
\orcid{0000-0001-5009-2327}
\affiliation{%
  \institution{Karlsruhe Institute of Technology}
  \city{Karlsruhe}
  \country{Germany}
}

\renewcommand{\shortauthors}{Jonas Hummel et al.}

\begin{abstract}
We present \systemname{}, a closed-loop earable system for stress-adaptive meditation that integrates in-ear physiological sensing with personalized, real-time intervention. Leveraging OpenEarable 2.0’s multimodal sensing, \systemname{} continuously monitors physiological signals and detects elevated stress from heart rate and heart rate variability. Upon detection, the system initiates a personalized guided meditation generated by an LLM and adapted in real time to the user’s stress state. The demo offers a hands-on experience of stress-adaptive meditation in two modes: a biosignal-adaptive meditation with optional stress induction to illustrate closed-loop adaptation, and a meditation-only mode focusing on \systemname{}’s generative personalization capabilities. The demo highlights how in-ear sensing, closed-loop adaptation, and personalized generative meditation can be integrated into an earable system for real-time stress support in demanding office work contexts.
\end{abstract}




\maketitle

\begin{figure*}[t]
    \centering
    \includegraphics[width=1\linewidth]{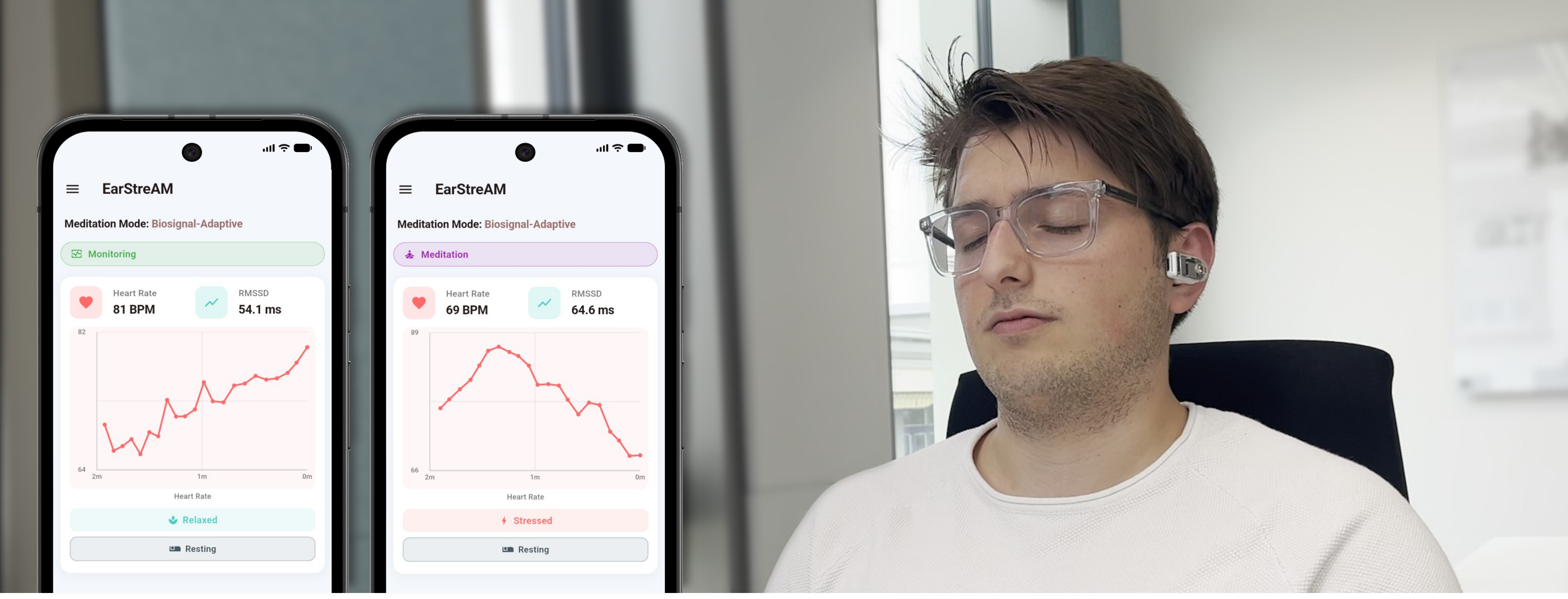} 
    \caption{\systemname{} helps users cope with stress in office settings by combining real-time sensing from OpenEarable 2.0~\cite{roddiger_openearable_2025} with a smartphone application. The system continuously tracks heart rate and heart rate variability as stress proxies and starts a personalized LLM-generated meditation when they exceed a predefined threshold, continuing until stress subsides.}
    \Description{\systemname{} helps users cope with stress in office settings by combining real-time sensing from OpenEarable 2.0~\cite{roddiger_openearable_2025} with a smartphone application. The system continuously tracks heart rate and heart rate variability as stress proxies and starts a personalized LLM-generated meditation when they exceed a predefined threshold, continuing until stress subsides. The figure shows a relaxed person wearing OpenEarable 2.0. The earable is connected to a smartphone, which is also shown. On this smartphone, the \systemname{} application is running, showing how heart rate is going up, triggering a meditation, which then leads to heart rate going down again.}
    \label{fig:teaser}
\end{figure*}

\section{Introduction}

Stress is a pervasive aspect of everyday life that affects people when following the news \cite{kesner_impact_2025}, navigating traffic \cite{antoun_acute_2017}, or in workplace settings \cite{lazarus_psychological_1995}. In the latter context, elevated stress has been associated with reduced productivity, lower job satisfaction, and increased turnover intentions \cite{bhat_examination_2023, bui_workplace_2021, gerhardt_how_2021}. To mitigate these effects, prior work has explored the effectiveness of brief meditation exercises for stress reduction that can be integrated into daily workplace routines \cite{woods-giscombe_workplace_2022}. However, many such approaches are not adaptive to a user’s momentary physiological state and therefore provide limited support when it is needed most.

To this end, we introduce \systemname{} (Earable-based Stress-Adaptive Meditation), a closed-loop earable system for personalized
stress-adaptive meditation. It combines OpenEarable 2.0 \cite{roddiger_openearable_2025} with a companion smartphone application to continuously capture physiological signals from in-ear photoplethysmography (PPG) and infer stress-related changes. When elevated stress is detected, the system initiates a personalized guided meditation generated by a large language model (LLM) that terminates once relaxation is detected. The system builds on StreAM \cite{bennardo_stream_2026}, an early-stage research prototype, and advances it toward a self-contained and practically deployable system for everyday workplace use.

Our demo gives attendees a hands-on experience of stress-adap-tive meditation in two modes. In the first mode, participants may undergo a brief stress induction that triggers \systemname{}’s adaptive response in an accelerated demonstration setting. In the second mode, they can directly explore the personalized meditation experience without prior stress induction. Through this demo, attendees can experience how in-ear sensing, closed-loop adaptation, and personalized, LLM-generated guided meditation can be combined into an interactive system for everyday stress support at the workplace.

\section{Related Work}


\subsection{Biosignal-Adaptive Stress Interventions}

Biosignal-based stress detection mechanisms have been widely explored \cite{giannakakis_review_2022, taskasaplidis_review_2024}, with PPG-derived heart rate features being among the most widely used stress operationalization constructs in wearable settings \cite{bolpagni_personalized_2024}. In contrast, systems that go beyond detection and implement closed-loop, biosignal-adaptive stress interventions remain comparatively underexplored \cite{jimenez-ocana_systematic_2023}. Existing work in this area has primarily focused on biofeedback-based approaches, including resonant breathing interventions guided by heart rate for stress reduction in workplace contexts \cite{purwandini_sutarto_resonant_2012}, as well as interreality systems that adapt virtual environments in response to physiological signals \cite{pallavicini_interreality_2013}. With regard to earables, recent work has demonstrated earable-based stress detection \cite{rahman_detecting_2022} and conceptualized the potential of in-ear sensing for closed-loop adaptive interventions \cite{xu_earable_2025}. However, to the best of our knowledge, no earable system has yet been demonstrated that combines continuous physiological stress detection with automatic initiation and real-time adaptation of interventions within a single device.

\subsection{LLM-Based Adaptive Meditation Systems}

Recent work has begun exploring the use of LLMs for personalized stress management and meditation support \cite{nguyen_ai-driven_2024, wu_mindfulagents_2026}. This direction was further advanced by StreAM \cite{bennardo_stream_2026}, which combined personalized, LLM-generated meditation guidance with physiological stress sensing in a biosignal-adaptive meditation system for workplace environments. StreAM demonstrated the feasibility and perceived usefulness of adaptive meditation experiences driven by physiological data. Building on this work, we leverage the sensing and audio capabilities of OpenEarable 2.0 to realize a fully closed-loop intervention that relies solely on an earable device and an accompanying smartphone, thereby improving suitability for everyday use. In addition, our system introduces refinements targeting robustness, personalization, and usability.

\section{\systemname{}: A Closed-Loop Earable System for Personalized
Stress-Adaptive Meditation}

\begin{figure*}[t]
    \centering
    \includegraphics[width=1\linewidth]{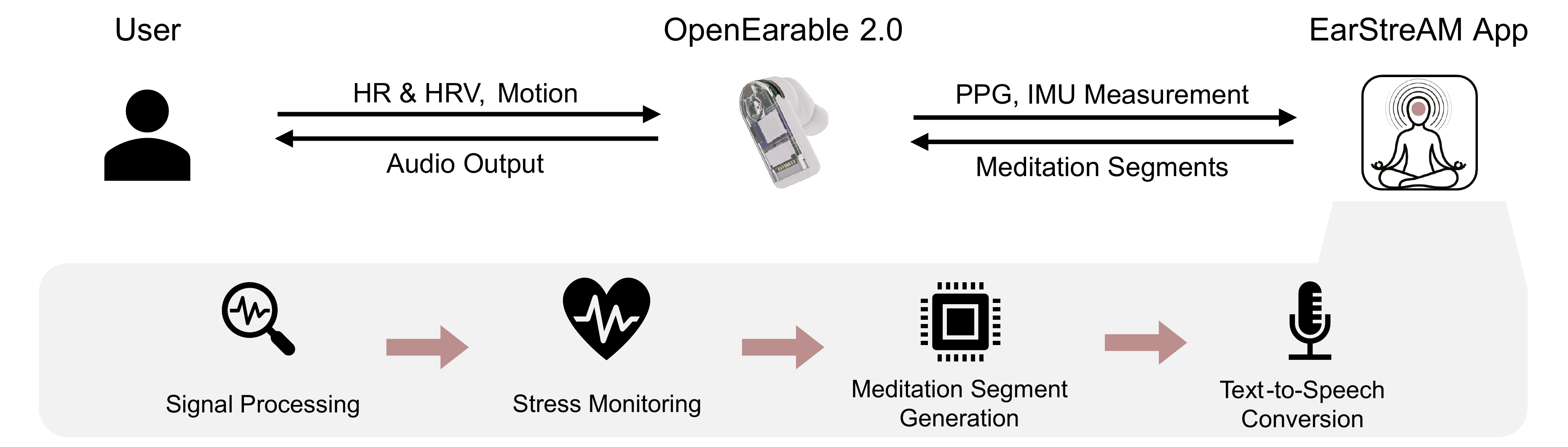} 
    \caption{\systemname{} constitutes a closed-loop biosignal-adaptive system. User biosignals are captured by OpenEarable 2.0 via its PPG and IMU sensors and transmitted to the mobile application via BLE for signal processing and continuous stress monitoring. When stress levels exceed a predefined threshold, the system generates meditation segments, converts them to speech via text-to-speech synthesis, and streams the resulting audio back to OpenEarable 2.0 for playback to the user. The user’s physiological response is then continuously monitored, thereby closing the adaptive feedback loop.}
    \Description{\systemname{} constitutes a closed-loop biosignal-adaptive system. User biosignals are captured by OpenEarable 2.0 via its PPG and IMU sensors and transmitted to the mobile application via BLE for signal processing and continuous stress monitoring. When stress levels exceed a predefined threshold, the system generates meditation segments, converts them to speech via text-to-speech synthesis, and streams the resulting audio back to OpenEarable 2.0 for playback to the user. The user’s physiological response is then continuously monitored, thereby closing the adaptive feedback loop.}
    \label{fig: earstream_structure_diagram}
\end{figure*}

\systemname{} combines OpenEarable 2.0 as both an earphone and sensing device with a companion smartphone application. The system continuously monitors a user’s physiological state using heart rate (HR) and heart rate variability (HRV) derived from in-ear PPG. Upon detecting elevated stress, the system initiates an LLM-generated personalized guided meditation. The meditation dynamically adapts to the user’s physiological response and terminates once stress levels return to baseline. In addition, \systemname{} provides a mode for experiencing the personalized meditation only. The overall system architecture is presented in \autoref{fig: earstream_structure_diagram}.



\subsection{Signal Processing Pipeline}

\systemname{} derives cardiovascular metrics from in-ear PPG sampled at 200 Hz, minimizing HRV measurement error \cite{rahman_detecting_2022}, with a 100~Hz accelerometer for motion referencing. A pre-session seal check \cite{kuttner_earresp-ans_2026} verifies sensor contact. Raw samples undergo adaptive artifact removal with channel selection across green/red/IR, ambient-light cancellation, and accelerometer-referenced motion cancellation with motion-aware outlier suppression before a DC-blocking high-pass filter, second-order IIR band-pass filter (0.45 - 5.5 Hz), and adaptive envelope normalizer. Peaks are detected on a 10 s sliding window using an adaptive amplitude threshold and minimum inter-peak spacing to avoid double-counting beats; RR intervals are plausibility-validated (250 - 2000 ms) and median-outlier-cleaned. A composite signal-quality score (waveform quality, motion, peak yield, RR regularity) gates emission during poor contact. HR is 60/median RR (s), Kalman-smoothed ($Q=0.02, R=5.0$). RR intervals accumulate in a rolling buffer of up to 5 min (minimum 1 min) from which RMSSD, reflecting parasympathetic activity \cite{shaffer_overview_2017}, is recomputed alongside every HR update. HR is emitted every 3 s after a 10 s warm-up; RMSSD follows once available.

\subsection{Context-Aware Stress Detection}

Stress detection in \systemname{} accounts for inter-individual variability in cardiovascular baselines through a hybrid estimation approach. Upon initialization, baseline values are derived from demographic information (age, gender, and fitness level) based on established population-level findings \cite{nunan_quantitative_2010, tegegne_determinants_2018, task_force_1996}. During the first five minutes of monitoring, the system progressively refines this baseline using real-time physiological measurements, resulting in a predominantly personalized reference that combines prior knowledge with observed data (30\% prior and 70\% data-driven). 

Stress is inferred using threshold-based comparisons between current physiological values and the individualized baseline. By default, elevated stress is detected when both HR increases and HRV decreases beyond a predefined relative threshold. Consistent with prior work \cite{castaldo_acute_2015}, this threshold is set to 10\% by default but can be adapted based on individual preferences. Stress detection operates as a continuous background process, reevaluating the user’s physiological state every 3 s using rolling HR and HRV estimates derived from the continuously updated RR interval stream. A meditation is triggered when sustained stress is detected. During an active session, the system reassesses the user’s state after each meditation segment to determine whether stress levels have decreased. For this purpose, HR is used as the sole recovery indicator, as HRV typically responds more slowly \cite{shaffer_overview_2017}. The meditation is terminated once HR returns to, or falls below, the individualized baseline.

Designed for office settings, the system incorporates an activity classification heuristic to distinguish stress-related physiological changes from those caused by movement. Activity is estimated using triaxial accelerometer and gyroscope data, based on a 3-second moving window of accelerometer deviation from gravitational acceleration and angular velocity. Periods of rest are identified when both accelerometer deviation ($<0.2$g) and gyroscope magnitude ($<300$°/s) remain low. Higher values are categorized as light ($0.2-0.4$g, $300-600$°/s), moderate ($0.4-0.8$g, $600-1200$°/s), vigorous ($0.8-2.0$g, $1200-3000$°/s), or intense ($>2.0$g, $>3000$°/s) activity. Meditation is not initiated if recent activity within a five-minute window indicates sustained movement, defined as at least 40\% light, 20\% moderate, 5\% vigorous, or 2\% intense activity.

\begin{figure}[b]
    \centering
    \includegraphics[width=1\linewidth]{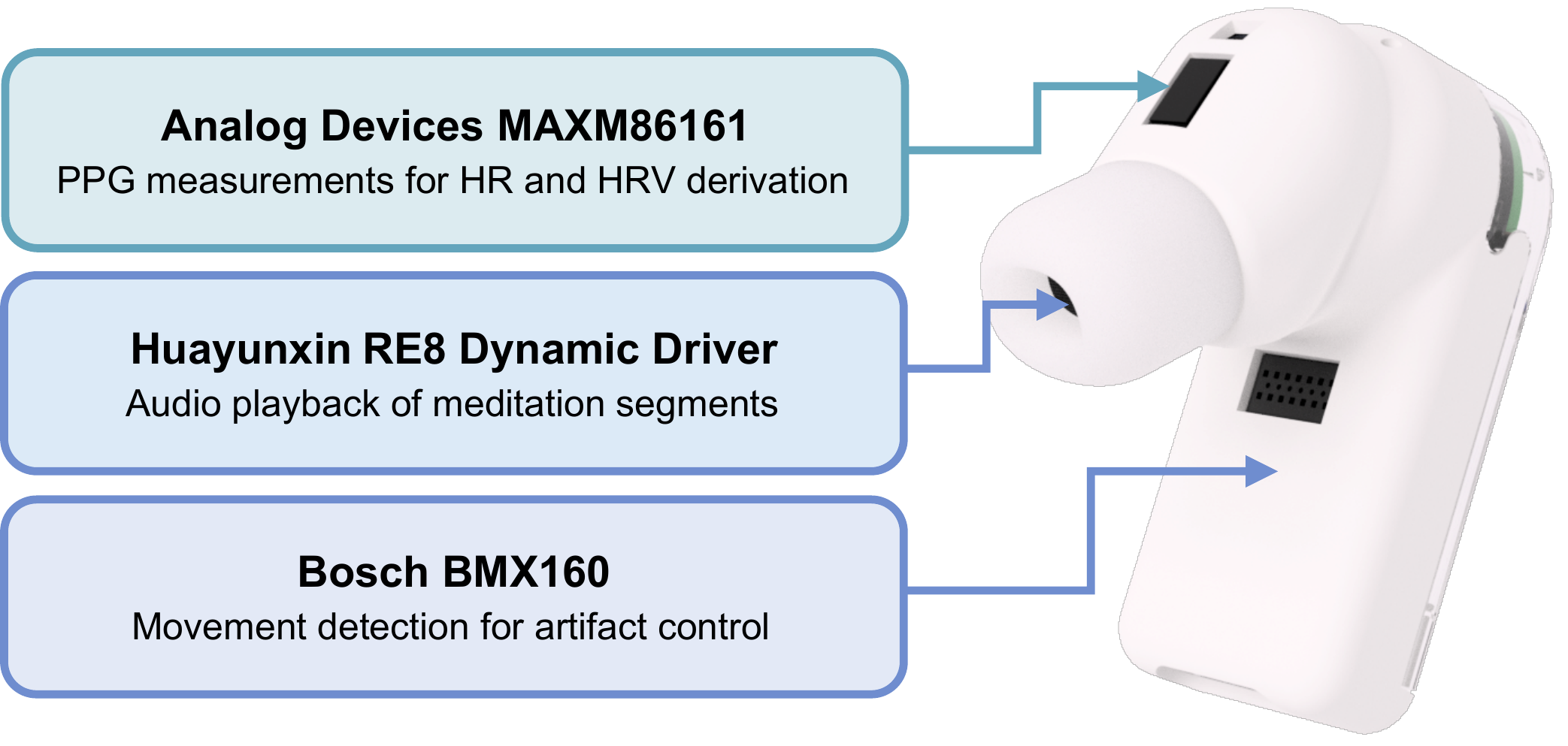} 
    \caption{\systemname{} is implemented using OpenEarable 2.0 \cite{roddiger_openearable_2025} and its multimodal sensing and actuation capabilities.}
    \Description{\systemname{} is implemented using OpenEarable 2.0 \cite{roddiger_openearable_2025} and its multimodal sensing and actuation capabilities. The used modalities are: 1. Analog Devices MAXM86161 (PPG measurements for HR and HRV derivation); 2. Huayunxin RE8 Dynamic Driver (Audio playback of meditation segments); 3. Bosch BMX160 (Movement detection for artifact control)}
    \label{fig: earable sensors}
\end{figure}

\begin{figure*}[t]
    \centering
    \includegraphics[width=1\linewidth]{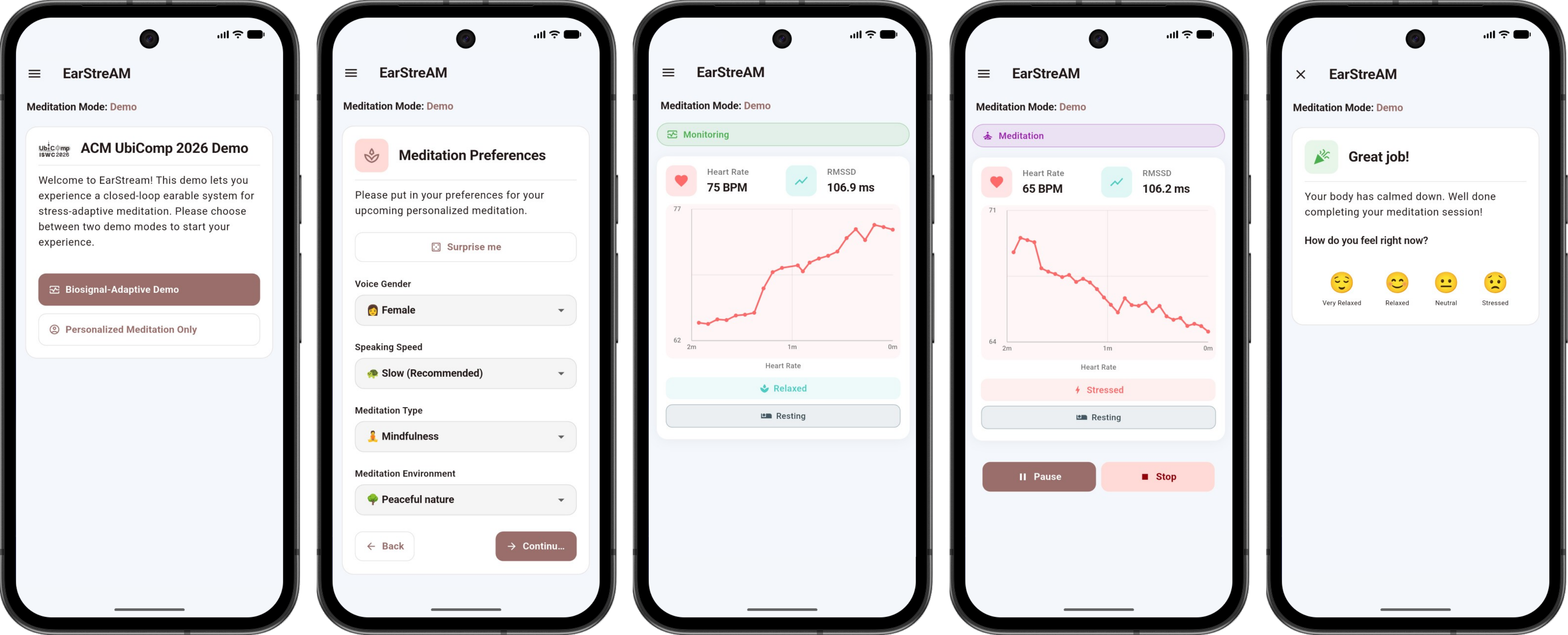} 
    \caption{Attendees choose between a biosignal-adaptive and a personalized meditation-only mode, and specify their preferences. While the personalized mode starts the generated meditation immediately, the biosignal-adaptive mode first enters a monitoring phase. Once elevated stress is detected, a personalized meditation is generated and played until stress decreases.}
    \Description{The figure shows the different phases of the demo attendees will experience. 1. Start of \systemname{} via a special demo mode. 2. Demo attendees can choose between a biosignal-adaptive and a personalized meditation-only mode, and specify their meditation preferences. 3. While the personalized mode starts the generated meditation immediately, the biosignal-adaptive mode first enters a monitoring phase. 4. Once elevated stress is detected, a personalized meditation is generated and played until stress levels decrease. 5. A final screen indicates when the meditation has ended successfully.}
    \label{fig: screenshot_walkthrough}
\end{figure*}








\subsection{Personalized Meditation Generation}

\systemname{} generates guided meditation tailored to user preferences and current physiological state. Users can configure key meditation aspects, including voice characteristics (gender, speaking rate), thematic elements (e.g., ocean), and meditation style (e.g., body scan). Users may also provide free-form preferences that are incorporated into the generation process. Together with the current physiological measurements, these inputs are integrated into a prompt passed to the LLM to generate personalized meditation in approximately 280-word segments. In our implementation, we employ DeepSeek’s V4 Pro text-generation model. The generated content is subsequently converted into audio using ElevenLabs Flash v2.5 or Tencent Cloud TTS with meditation-optimized voice settings. \systemname{} thereby enables real-time personalization at scale, generating meditation content that adapts to both the user's physiological state and personal preferences within each session.

\section{Demonstration}

We demonstrate \systemname{} through an interactive setup that lets attendees experience real-time physiological stress monitoring and adaptive guided meditation. Participants receive OpenEarable 2.0 devices and a companion smartphone, enabling continuous in-ear physiological sensing and adaptive audio feedback.

The demo offers two modes. In the first mode, participants experience a time-condensed \systemname{} cycle showcasing real-time stress detection and adaptive intervention. The system performs a shortened 30 s baseline measurement~\cite{shaffer_overview_2017} based solely on physiological data, followed by ultra-sensitive stress detection (threshold set to 3\%) and shorter meditation segments of approximately 30 s. To elicit a measurable stress response within the limited demo duration, participants may undergo a brief stress induction inspired by the Maastricht Acute Stress Test \cite{shilton_maastricht_2017}, combining cold water exposure with a timed mental arithmetic task. In the second mode, participants explore personalized guided meditation without biosignal-based stress-adaptation. This mode emphasizes the system’s generative personalization capabilities, allowing attendees to engage with individualized meditation content in a more relaxed setting. For privacy, physiological processing remains local to the smartphone, while prompts sent to the LLM and TTS services exclude personal and physiological information. Together, the two modes illustrate how continuous in-ear sensing, closed-loop adaptation, and generative guidance are integrated into an earable system for real-time stress support. They further show how such interventions can provide moments of relief in demanding professional contexts, as reflected in the conference environment.



\section{Conclusion}

We presented \systemname{}, a self-contained earable system for stress-adaptive meditation in everyday workplace environments that combines in-ear physiological sensing, personalized LLM-generated guidance, and audio feedback on OpenEarable 2.0. Through our interactive demo, attendees can experience how continuous in-ear sensing and real-time adaptation enable personalized stress interventions in demanding contexts. We hope this demonstration motivates future research on closed-loop earable interventions and contributes to the development of context-aware stress coping technologies that provide timely and individualized support.














\begin{acks}
Funded by the Deutsche Forschungsgemeinschaft (DFG, German Research Foundation) – GRK2739/2 – Project Nr. 447089431 – Research Training Group: KD2School – Designing Biosignal-Adaptive Systems for Decision-Making Processes
\end{acks}

\bibliographystyle{ACM-Reference-Format}
\balance
\bibliography{references}


\end{document}